\documentclass[a4paper,fleqn,usenatbib,useAMS]{mnras}

\usepackage{graphicx}	% Including figure files
\usepackage{amsmath}	% Advanced maths commands
\usepackage{amssymb}	% Extra maths symbols
\usepackage{multicol}        % Multi-column entries in tables

\usepackage{bm}		% Bold maths symbols, including upright Greek
\usepackage{pdflscape}	% Landscape pages
\usepackage{xcolor}

\newcommand\bb[1]{\mbox{\boldmath{$#1$}}}
\newcommand\del{\bb{\nabla}}
\newcommand\bcdot{\bb{\cdot}}

\newcommand\elec{E}

\usepackage[T1]{fontenc}
\usepackage{ae,aecompl}

\usepackage{newtxtext,newtxmath}
\title[Magnetogenesis at Cosmic Dawn]{Magnetogenesis at Cosmic Dawn: Tracing the Origins of Cosmic Magnetic Fields}

\author[H. Katz \& S. Martin-Alvarez et al.]{Harley Katz$^{1}$\thanks{Contact e-mail: \href{mailto:harley.katz@physics.ox.ac.uk}{harley.katz@physics.ox.ac.uk}}, Sergio Martin-Alvarez$^{1}$\thanks{Co-First Author, \href{mailto:sergio.martin@physics.ox.ac.uk}{sergio.martin@physics.ox.ac.uk}}, Julien Devriendt$^{1,2}$, Adrianne Slyz$^1$, \newauthor \& Taysun Kimm$^3$
\\
$^{1}$Astrophysics, University of Oxford, Denys Wilkinson Building, Keble Road, Oxford OX1 3RH, UK\\
$^{2}$Universit\' e de Lyon, Universit\' e Lyon 1, ENS de Lyon, CNRS, Centre de Recherche Astrophysique de Lyon UMR5574, F-69230 Saint-Genis-Laval, France\\
$^{3}$Department of Astronomy, Yonsei University, 50 Yonsei-ro, Seodaemun-gu, Seoul 03722, Republic of Korea\\
}

\date{\today}

\pubyear{2018}

\begin{document}
\label{firstpage}
\pagerange{\pageref{firstpage}--\pageref{lastpage}}
\maketitle

\begin{abstract}
Despite their ubiquity, the origin of cosmic magnetic fields remains unknown.  Various mechanisms have been proposed for their existence including primordial fields generated by inflation, or amplification and injection by compact astrophysical objects.  Separating the potential impact of each magnetogenesis scenario on the magnitude and orientation of the magnetic field and their impact on gas dynamics may give insight into the physics that magnetised our Universe.  In this work, we demonstrate that because the induction equation and solenoidal constraint are linear with $B$, the contribution from different sources of magnetic field can be separated in cosmological magnetohydrodynamics simulations and their evolution and influence on the gas dynamics can be tracked.  Exploiting this property, we develop a magnetic field tracer algorithm for cosmological simulations that can track the origin and evolution of different components of the magnetic field.  We present a suite of cosmological magnetohydrodynamical {\small RAMSES} simulations that employ this algorithm where the primordial field strength is varied to determine the contributions of the primordial and supernovae-injected magnetic fields to the total magnetic energy as a function of time and spatial location.  We find that, for our specific model, the supernova-injected fields rarely penetrate far from haloes, despite often dominating the total magnetic energy in the simulations. The magnetic energy density from the supernova-injected field scales with density with a power-law slope steeper than 4/3 and often dominates the total magnetic energy inside of haloes.  However, the star formation rates in our simulations are not affected by the presence of magnetic fields, for the ranges of primordial field strengths examined.  These simulations represent a first demonstration of the magnetic field tracer algorithm which we suggest will be an important tool for future cosmological MHD simulations.
\end{abstract}


%%%%%%%%%%%%%%%%%%%%%%%%%%%%%%%%%%%%%%%%%%%%%%%%%%

%%%%%%%%%%%%%%%%% BODY OF PAPER %%%%%%%%%%%%%%%%%%

\section{Introduction}
Magnetic fields are ubiquitous throughout our Universe \citep{Widrow2002}. They have been observed in the most compact objects such as stars \citep{Reiners2012} and black holes \citep{Johnson2015}, on galaxy scales such as in our own Milky Way \citep{Davis1951,Mulcahy2014}, in galaxy groups and clusters \citep{Large1959,Carilli2002,Govoni2004}, and finally in the intergalactic medium \citep{Kim1989,Kronberg1994,Grasso2001}. Regardless of their recognised presence and importance in all these environments, many details regarding their evolution and particularly their origin remain unknown. 

Due to their complexity, numerical simulations have become the primary tool for improving our understanding of cosmic magnetic fields. A significant body of literature already exists regarding the use of numerical simulations to model magnetic fields on small, sub-galactic \citep{Hennebelle2014,Evirgen2017,Gomez2018}, galaxy \citep{Wang2009,Pakmor2017}, galaxy cluster \citep{Dolag1999,Dubois2008,Marinacci2015,Vazza2015}, and cosmological scales \citep{Ryu08,Marinacci2017,Alves-Batista2017}. These similar numerical magnetohydrodynamical (MHD) simulations are also of paramount importance for the study of astrophysical plasma processes \citep{Schekochihin2004,Federrath2016}.

Upper and lower bounds on cosmic magnetic field strength exist from a variety of probes. An upper limit of $B\lesssim 10^{-9}$G is placed by studying CMB B-mode perturbations \citep{Planck2016,Pogosian2018}, and lower limits of $B\gtrsim 10^{-17}$G are available from gamma-rays particle cascades \citep{Neronov2010}. Note that the validity of these lower limits has been debated \citep[and subsequent studies]{Broderick2012}.

In the interstellar medium (ISM) of galaxies, the energy contained in the magnetic field is observed to be in rough equipartition ($B\sim \mu$G) with the thermal and turbulent energy of the galaxy, even at high redshifts \citep{Bernet2008}. Current theoretical studies tend to favour two different models to attain these values above a weak primordial field: either dynamo amplification \citep{Kulsrud2008,Pakmor2014,Rieder2016,Martin-Alvarez18} or by magnetised feedback from stars \citep{Beck2013,Butsky2017} or black holes \citep{Vazza2017}. These two scenarios were already discussed by \citet{Rees1987}, illustrating the longevity of this question. While all appear individually sufficient to produce realistic magnetisation in galaxies, it is difficult to disentangle the contribution from primordial magnetic fields, dynamo amplification, and compact astrophysical sources when all operate simultaneously and to determine which of these mechanisms dominate the magnetic fields in galaxies.  

A similar difficulty persists for galaxy clusters, where debated origins of magnetic fields range from primordial or plasma dynamo processes to magnetisation through Active Galactic Nuclei (AGN) feedback. Extensive numerical work addressing possible mechanisms exists \citep{Ryu08,Vazza2018}, and of particular importance are the observational predictions differentiating each scenario \citep{Donnert2009,Vazza2017}.

A last meaningful unknown is the provenance of cosmic magnetic fields. Within the aforementioned observational limits, magnetic fields in the IGM and on cosmological scales are yet to be understood \citep{Widrow2002}.  Although many theoretical possibilities exist, the origins of primordial magnetic fields during or after inflation have yet to be determined \citep{Kandus2011,Subramanian2016}.  Furthermore, the different scenarios are not mutually exclusive and how the different primordial magnetic fields interact with other magnetic fields such as those escaping from galaxies \citep{Dubois2010} and galaxy clusters \citep{Sutter2012} to produce the total cosmic magnetic field remains unknown. Equally, if and how these primordial magnetic fields could be influencing smaller systems is uncertain and difficult to constrain due to the plethora of possible models \citep{Marinacci2016}.

All of the described problems share a common characteristic: it is extremely complicated to separate contributions from different physical processes to the resulting magnetic fields. As a consequence, understanding the different possible mechanisms for magnetic field generation and amplification and significance to the total magnetic field currently cannot be studied in a methodical manner. This has been one of the major obstacles for progress in this field. Since the origin of magnetic fields can likely be traced to a variety of different sources, both of primordial and astrophysical nature, ideally one would be able to differentiate the contribution from each source to the total B-field and total magnetic pressure/energy at every place and time in the Universe. Previous works on this topic have generally taken the approach of turning on and off the different physics in order to understand how structure, star formation, and magnetisation evolves differently under changing scenarios \citep{Beck2013,Vazza2017,Martin-Alvarez18,Steinwandel2018}. However, the different sources of B-fields are not necessarily mutually exclusive and each may be dynamically important and affect either the generation or evolution of the other. Thus we aim to develop a method where the total B-field in a simulation can be affected by a variety of sources, yet the contribution from each can be tracked in order to better understand different scenarios of magnetogenesis. In what follows, we develop a new algorithm to trace the contribution of different classes of sources to the total B-field in cosmological magneto-hydrodynamic simulations. 

The paper is organised as follows. Section~\ref{s:Algorithm} describes the theoretical foundations and our implementation in the {\sc RAMSES} code. Section~\ref{s:Simulations} describes the different simulations we use to demonstrate the potential of our new algorithm.  We describe the results in Section~\ref{s:Results} with caveats listed in Section~\ref{s:Caveats}. A discussion can be found in Section~\ref{s:Conclusions}.

\section{Magnetic Field Tracers}
\label{s:Algorithm}

We take advantage of the linearity of the induction equation and the solenoidal constraint for the evolution of the magnetic field to develop a method that separately follows the individual contributions to the total magnetic field from a variety of sources in cosmological simulations.  Note that our implementation is designed for grid-based codes.

\begin{figure*}
\centerline{\includegraphics[scale=1]{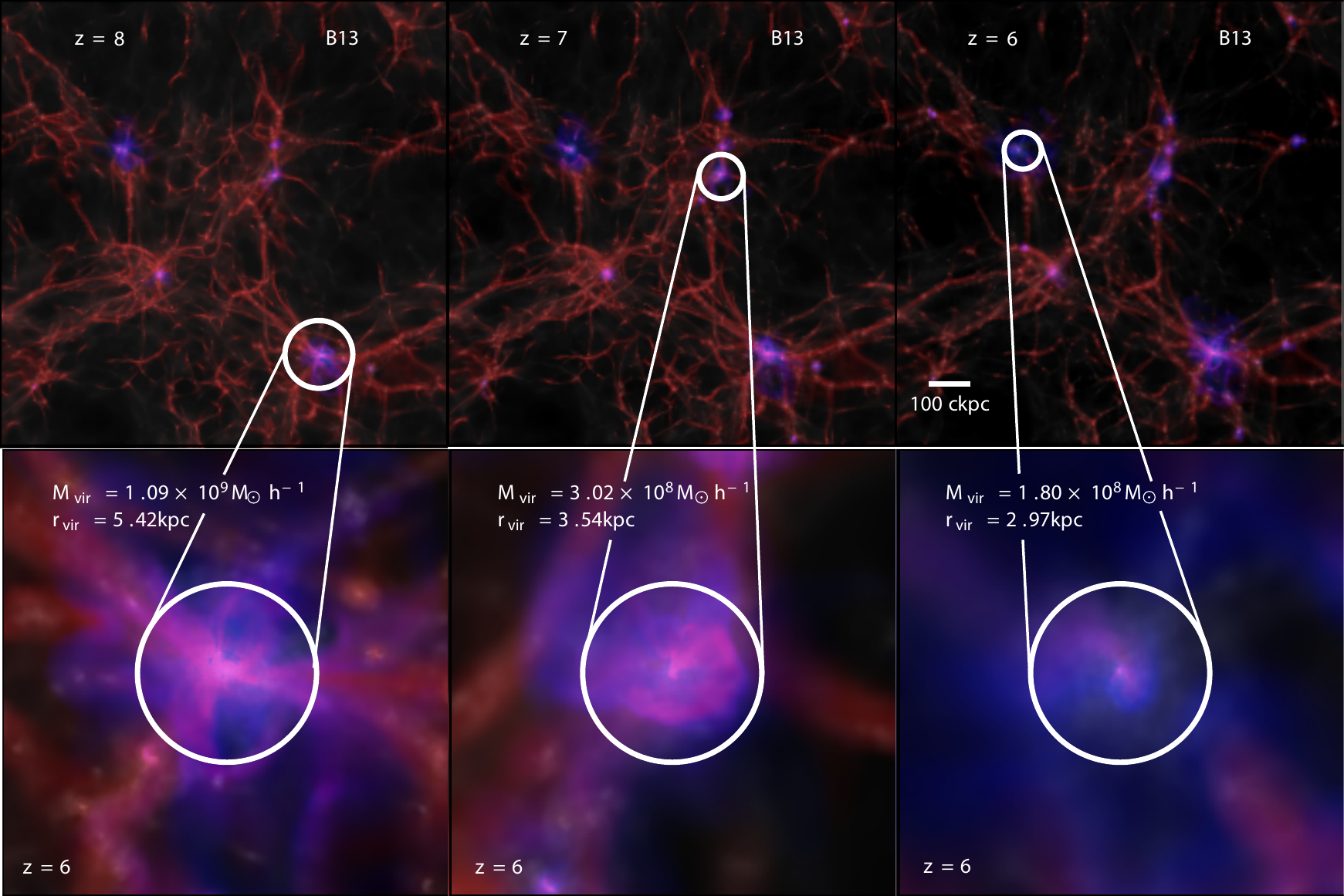}}
\caption{(Top) Maps of the magnetic field strength across the full simulation volume for three different redshifts from our B13 simulation. The red regions represent the magnitude of the primordial magnetic field while the blue regions indicate the magnitude of the magnetic field injected during SNe explosions.  The dark matter column density is under-laid in grey-scale and the images show a projection along the $z$-direction of the simulated box. (Bottom)  Maps of the magnetic field around the three most massive haloes at $z=6$.  The virial radii of the haloes are indicated with the white circles and scaled to be the same size in all plots. }
\label{MHDrgb}
\end{figure*}

The equations for ideal MHD, written in conservative form are:
\begin{eqnarray}
\frac{\partial \rho}{\partial t} + \del \bcdot (\rho \bb{v})  =  0 , \\
\frac{\partial \rho \bb{v}}{\partial t} + \del \bcdot (\rho\bb{v}\bb{v} - \bb{B}\bb{B}) + \del P_{\rm tot} =0 , \\
\frac{\partial E}{\partial t} + \del \bcdot \left[ (E+P_{\rm tot})\bb{v}-\bb{B}(\bb{B} \bcdot \bb{v}) \right] =0 , \\
\frac{\partial \bf{B}}{\partial t}-\bf{\del\times(\bf{v}\times\bf{B})}=0,
\label{mhd equations}
\end{eqnarray}
where $\rho$ is the gas density, $\bb{v}$ is the fluid velocity, $\bb{B}$ is the magnetic field, $P_{\rm tot}$ is the total pressure (thermal and magnetic), and $E$ is the total energy (thermal, kinetic, and magnetic).  In order to isolate the contribution of different sources to the B-field we allow the total B-field ($\bb{B}$) to be written as
\begin{equation}
    \bb{B}=\sum_m^{\rm N_{source}}\bb{B}_{t_m},
    \label{eq:Decomposition}
\end{equation}
where ${\rm N_{source}}$ is the total number of different mechanisms that generate a B-field. Figure~\ref{MHDrgb} presents an example where a primordial magnetic field (red), and one that is injected via supernova (blue) are traced. $\bb{B}_{t_m}$ can be thought of as the B-field generated by an individual source that evolves without knowledge of the other $\bb{B}_{t_m}$'s except for the response of the fluid to the presence of the total B-field in the simulation. In other words, the dynamics of the fluid in the simulation only respond to the total B-field, $\bb{B}$, which then has a dynamical effect on the evolution of each $\bb{B}_{t_m}$; however, each $\bb{B}_{t_m}$ is not aware of the presence of any other $\bb{B}_{t_m}$ when the induction equation is solved or the electromotive forces (EMFs) are calculated. The tracer algorithm allows us to trace the amplification and reduction of each tracer magnetic field separately throughout cosmic time. This is only possible because the induction equation (Equation~4) and the solenoidal constraint are linear in $\bb{B}$. 

Beginning with the induction equation we can expand it in three dimensions into the following form,

\begin{equation}
    \begin{bmatrix}
    \frac{\partial B_x}{\partial t}\\
    \frac{\partial B_y}{\partial t}\\
    \frac{\partial B_z}{\partial t}
    \end{bmatrix}
-
    \begin{bmatrix}
    \frac{\partial}{\partial y}(v_xB_y-v_yB_x)-\frac{\partial}{\partial z}(v_zB_x-v_xB_z) \\
    \frac{\partial}{\partial z}(v_yB_z-v_zB_y)-\frac{\partial}{\partial x}(v_xB_y-v_yB_x) \\
    \frac{\partial}{\partial x}(v_zB_x-v_xB_z)-\frac{\partial}{\partial y}(v_yB_z-v_zB_y)
    \end{bmatrix}
=
    \begin{bmatrix}
    0\\
    0\\
    0
    \end{bmatrix}.
\end{equation}

By placing a subscript of $t_m$ on each of the $B$ components in the previous equation, we can obtain the advection equation for each of the individual tracer fields. In more detail, if we consider only the change in the $B_x$ component, we have
\begin{equation}
    \begin{split}
    \frac{\partial B_x}{\partial t} = \frac{\partial\sum_m^{\rm N_{source}}B_{x,t_m}}{\partial t} = \\ \frac{\partial}{\partial y}\left(v_x\sum_m^{\rm N_{source}}B_{y,t_m}-v_y\sum_m^{\rm N_{source}}B_{x,t_m}\right)\\ -\frac{\partial}{\partial z}\left(v_z\sum_m^{\rm N_{source}}B_{x,t_m}-v_x\sum_m^{\rm N_{source}}B_{z,t_m}\right).
    \end{split}
\end{equation}

Considering the simple case of only two tracer groups, 
\begin{equation}
    \begin{split}
    \frac{\partial B_x}{\partial t} = \frac{\partial (B_{x,t_1} + B_{x,t_2})}{\partial t} = \\ \frac{\partial}{\partial y}\left(v_x(B_{y,t_1}+B_{y,t_2})-v_y(B_{x,t_1}+B_{x,t_2})\right)\\ -\frac{\partial}{\partial z}\left(v_z(B_{x,t_1}+B_{x,t_2})-v_x(B_{z,t_1}+B_{z,t_2})\right)\\
    =\left[\frac{\partial}{\partial y}\left(v_xB_{y,t_1}-v_yB_{x,t_1}\right)
    - \frac{\partial}{\partial z}\left(v_zB_{x,t_1}-v_xB_{z,t_1}\right)\right]\\
    +\left[\frac{\partial}{\partial y}\left(v_xB_{y,t_2}-v_yB_{x,t_2}\right)
    - \frac{\partial}{\partial z}\left(v_zB_{x,t_2}-v_xB_{z,t_2}\right)\right]\\
    =\frac{\partial B_{x,t_1}}{\partial t} + \frac{\partial B_{x,t_2}}{\partial t}.
    \end{split}
\end{equation}
Hence the advection term in the induction equation satisfies the condition required for the total field to be separated into individual tracer groups. 

For our implementation, following \cite{Fromang2006}, the magnetic field is separated into a two-step solver, where we first compute the induction due to the generation of EMFs and we then advect the cell-centred magnetic field. The corresponding magnetic induction by the EMFs follows

\begin{equation}
\label{induction}
\begin{split}
\frac{B^{n+1}_{x,i-1/2,j,k}-B^n_{x,i-1/2,j,k}}{\Delta t}
-\frac{E^{n+1/2}_{z,i-1/2,j+1/2,k}-E^{n+1/2}_{z,i-1/2,j-1/2,k}}{\Delta y}\\
+ \frac{E^{n+1/2}_{y,i-1/2,j,k+1/2}-E^{n+1/2}_{y,i-1/2,j,k-1/2}}{\Delta z}=0 
  ,     
\end{split}
\end{equation}
where $n$ is the time step, $E$ is the time and edge-averaged EMF, and $i,j,k$ represent the edge or face perpendicular to the listed coordinate direction.  For instance, $B^{n}_{x,i-1/2,j,k}$ gives the value of the magnetic field of the left face in the $x-$ direction at time step $n$ while $E^{n}_{z,i-1/2,j+1/2,k}$ gives the EMF at the top edge of the left face in the $z-$direction of the same cell at the same time.

It is clear that the first term in Equation~\ref{induction} is linear in $B$ so that 
\begin{equation}
\begin{split}
    \frac{B^{n+1}_{x,i-1/2,j,k}-B^n_{x,i-1/2,j,k}}{\Delta t} =
    \sum_m^{\rm N_{source}}\frac{B^{n+1}_{t_m,x,i-1/2,j,k}-B^n_{t_i,x,i-1/2,j,k}}{\Delta t},
\end{split}
\end{equation}
where $B^{n+1}_{t_m,x,i-1/2,j,k}$ is now the value of the magnetic field at the time $n+1$ at the left face in the $x-$direction for the source $t_m$.  The second two terms have the exact same form as the first so as long as the time and edge-averaged EMFs are linear in $B$, the induction equation can be proven to be separable.  For the EMFs, we have (for one specific edge)
\begin{eqnarray}
E^{n+1/2}_{z,i-1/2,j-1/2,k} = & \nonumber \\
\frac{1}{\Delta t \Delta z} 
\int_{t_n}^{t_{n+1}} 
\int_{z_{k-1/2}}^{z_{k+1/2}} & 
\elec_z(x_{i-1/2},y_{j-1/2},z',t')dz'dt' \, ,
\end{eqnarray}
and thus
\begin{equation}
\label{emf2}
\begin{split}
E^{n+1/2}_{z,i-1/2,j-1/2,k} = \\
\frac{1}{\Delta t \Delta z} 
\int_{t_n}^{t_{n+1}} 
\int_{z_{k-1/2}}^{z_{k+1/2}}  
\sum_m^{\rm N_{source}}\elec_(t_m,z)(x_{i-1/2},y_{j-1/2},z',t')dz'dt'=\\
\frac{1}{\Delta t \Delta z} 
\sum_m^{\rm N_{source}}\int_{t_n}^{t_{n+1}} 
\int_{z_{k-1/2}}^{z_{k+1/2}} 
\elec_{t_m,z}(x_{i-1/2},y_{j-1/2},z',t')dz'dt',
\end{split}
\end{equation}
if 
\begin{equation}
\label{emf1}
    \elec_z(x_{i-1/2},y_{j-1/2},z,t) = \sum_m^{\rm N_{source}}\elec_{t_m,z}(x_{i-1/2},y_{j-1/2},z,t).
\end{equation}
Since,
\begin{equation}
E_{z,i-1/2,j-1/2,k}^{n}=\bar{v}_x\bar{B}_y-\bar{v}_y\bar{B}_x \, ,
\end{equation}
where
\begin{equation}
\begin{split}
\bar{v}_x &= \frac{1}{4} (v^n_{x,i,j,k}+v^n_{x,i-1,j,k}+v^n_{x,i,j-1,k}+v^n_{x,i-1,j-1,k}) \, , \\
\bar{v}_y &= \frac{1}{4} (v^n_{y,i,j,k}+v^n_{y,i-1,j,k}+v^n_{y,i,j-1,k}+v^n_{y,i-1,j-1,k}) \, , \\
\bar{B}_x &= \frac{1}{2} (B^n_{x,i-1/2,j,k}+B^n_{x,i-1/2,j-1,k}) \, , \\
\bar{B}_y &= \frac{1}{2} (B^n_{y,i,j-1/2,k}+B^n_{y,i-1,j-1/2,k})\, ,
\end{split}
\end{equation}
it is clear that neither $\bar{v}_x$ nor $\bar{v}_y$ are dependent on $B$ and both $\bar{B}_x$ and $\bar{B}_y$ are linear in $B$, thus Equation~\ref{emf1} holds, which ensures that Equation~\ref{emf2} is also true.  Hence the induction is entirely linear in $B$ and therefore separable into the different components, $ B_{t_m}$. As we solve the induction for each of the tracers following the same algorithm, the inducted contribution to each of the magnetic tracers is solenoidal by construction \citep[see][]{Fromang2006}. Advection of magnetic fields in {\sc RAMSES} is done through cell-centred fluxes. Each cell-centred flux is extracted from the corresponding two faces from the advecting cell, and added to the two corresponding faces of the advected cell. Therefore, this contribution is equally divergence-less. Reproducing this algorithm for the tracers, the computed fluxes for each tracer fulfil linearity, leading by construction to an equally solenoidal magnetic field.

In our implementation for this work, we will track the total field in the normal fashion as well as each individual tracer field.  This is to demonstrate that following the tracer fields individually exactly conserves all of the properties of the total magnetic field (see Appendix~\ref{appB}). The caveat of following all fields (as is done in the current work so that we can explicitly demonstrate the convergence properties) is that when reconstructing states on cell faces and edges, or refining a cell, slope limiters are generally used to ensure that the reconstruction is second-order total variation diminishing.  However, once a generic slope limiter (e.g. MinMod) is applied, it may no longer be the case that all magnetic field properties are perfectly conserved because the slopes may be limited by different amounts.  Thus in the current implementation, we have removed the slope-limiters for all the magnetic field quantities (note that hydrodynamic quantities are still slope-limited).  However, since one of the tracer fields can be reconstructed by knowing the total field as well as the other tracer fields, in future simulations we only need to follow ${\rm N_{source}-1}$ tracer fields.  By doing this, we can now apply the slope-limiter to the total field and hence obtain the same exact results as a simulation that does not include the tracers.  By definition, the reconstruction of the tracer field that is not explicitly followed will also be divergence-less (since the solenoidal constraint is linear in $B$), and the sum of all of the tracer fields will add up to the total.  Furthermore, this method is less computationally expensive as we can remove one of the fields with the caveat that the reconstruction is slightly less accurate than following every field individually.  Future work will apply this method to ensure that the simulation is second-order total variation diminishing.  

Note that this issue is not unique to our simulation.  In any scenario where, for instance, two or more quantities must exactly sum to a third, and each are individually slope-limited (e.g. in computational chemistry), there is no guarantee that after the calculation, the sum will be conserved.  Ordinarily this is not a major problem as the quantities can be rescaled but in the case of MHD, the solenoidal constraint prevents us from using this rescaling.  

\begin{figure*}
\centerline{\includegraphics[scale=1]{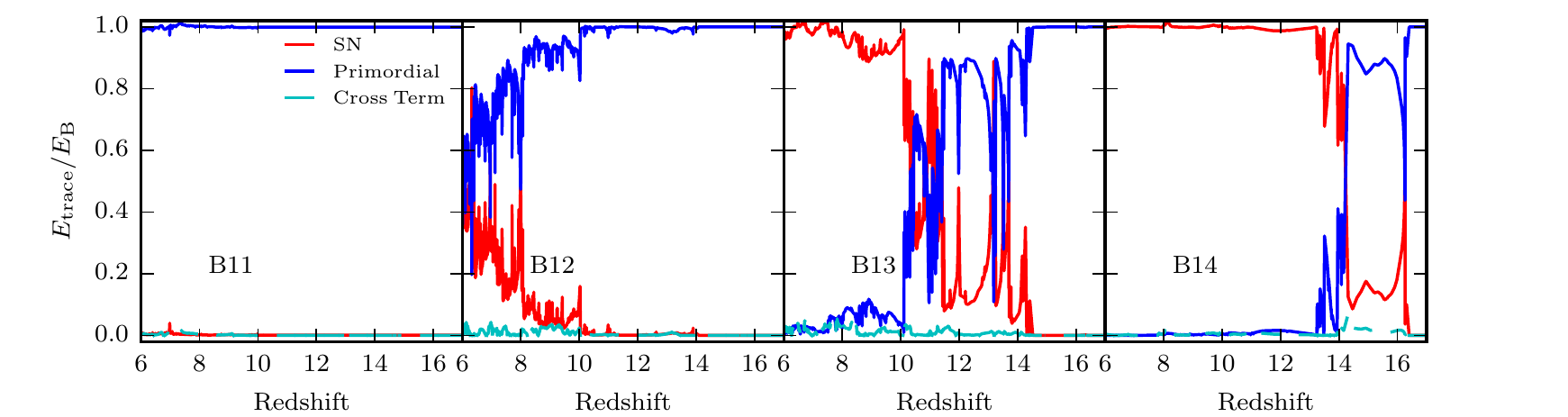}}
\caption{Fraction of the total magnetic energy in the box contained in the primordial field, the SN generated field, or the cross term between the two components as a function of redshift for each simulation.  The solid regions of the cyan line indicate redshifts at which the cross term is positive while the few dashed regions indicate that the cross term is negative. The simulations were designed to sample the parameter space where the primordial component dominates the energy at $z=6$ (B11) to where the SN component dominates at $z=6$ (B14). The B12 simulation exhibits approximately equipartition of the total magnetic energy at $z=6$.}
\label{enfrac}
\end{figure*}

\section{Numerical Simulations}
\label{s:Simulations}
Our MHD tracer algorithm has been implemented into {\small RAMSES} \citep{Teyssier2002}, an open-source, massively parallel, cosmological adaptive mesh refinement (AMR) code with a constrained transport \citep[CT,][]{Evans1988} implementation for ideal MHD \citep{Teyssier2006,Fromang2006}. We employ the HLLC Riemann solver \citep{Toro1994} to calculate the time-centred intercell fluxes and a MinMod slope limiter to reconstruct the cell-centred hydrodynamic properties at their faces (for non B-field quantities). We assume an adiabatic index of $\gamma=5/3$ (that of an ideal monatomic ideal gas) for the relation between gas pressure and internal energy. Particles (dark matter and stars) are projected onto the adaptive grid using cloud-in-cell interpolation to construct the density field needed to calculate gravitational forces. A multigrid scheme is then used to solve the Poisson equation on the grid \citep{Guillet2011}. 

Initial conditions were generated in a $1.25^3$Mpc$^3$ comoving box, discretised into a set of $128^3$ gas cells and dark matter particles using {\small MUSIC} \citep{Hahn2011}. The simulation was initialised at $z=150$, using the following cosmological parameters: $\Omega_m = 0.3175$, $\Omega_\Lambda = 0.6825$, $\Omega_b = 0.049$, $\sigma_8 = 0.83$, and Hubble constant $H_0 = 67.11 \text{km} \text{s}^{-1} \text{Mpc}^{-1}$, consistent with \cite{Planck2016}. All of our simulations include heating and cooling processes for the gas component. Metallicity dependent gas cooling rates are included at $T>10^4$K \citep{Sutherland1993} and for temperatures below $10^4$ K \citep{Rosen1995}. A UV background \citep{Haardt1996} models photo-ionisation instantaneously at redshift $z = 8.5$. We set an initial metallicity floor of $10^{-3.5} Z_{\odot}$ to mimic early enrichment by Population~III stars \citep{Wise2012}.

Throughout the simulation, cells are allowed to adaptively refine. We employ a quasi-Lagrangian approach whereby cells that contain at least integer multiples of eight times the initial dark matter mass or baryonic mass are allowed to refine into eight children cells. We set a fixed maximum refinement level of 14 which corresponds to a physical resolution of 10.9~pc at $z=6$.

When gas cells reach the maximum level of refinement, they are allowed to form star particles.  The local properties of the star forming clouds are expected to affect the efficiency of star formation \citep{Padoan2011,Hennebelle2011,Federrath2012}. Accordingly, our star formation prescription accounts for the properties of gas clouds in the simulations when forming stars. In our simulations, regions in the process of collapse are allowed to subsequently form stars only after this collapse cannot be further resolved. Our star formation prescription is based on a magneto-thermo-turbulent (MTT) Jeans length criterion. We define the MTT Jeans length
\begin{equation}
\lambda_\text{J,MTT} = \frac{\pi \sigma_V^2 + \sqrt{36 \pi c_\text{s,eff}^2 G {\Delta x}^2 \rho + \pi \sigma_V^4}}{6 G \rho {\Delta x}},
\end{equation}
where $G$ is the gravitational constant, $\sigma_V$ is the gas turbulent velocity, and ${\Delta x}$ is the length of a cell. In this equation, $c_\text{s,eff}^2$ is an effective sound speed defined to account for an isotropic small-scale contribution from magnetic pressure to the support of the gas.
\begin{equation}
    c_\text{s,eff} = c_\text{s} \sqrt{1 + \beta^{-1}},
\end{equation}
where $\beta = P_\text{thermal} / P_\text{mag}$ in a given cell. Thus, the generation of star particles is allowed only in cells where ${\Delta x} > \lambda_\text{J,MTT}$. In these cells, gas is converted into star particles employing a Schmidt law, with star formation rate
\begin{equation}
\dot{\rho}_\text{star} = \epsilon_\text{ff} \frac{\rho}{t_\text{ff}}.
\end{equation}
The free-fall time of the gas, $t_\text{ff}$, is defined as
\begin{equation}
t_\text{ff} = \sqrt{\frac{3\pi}{32 G \rho}},
\end{equation}
and $\epsilon_\text{ff}$ corresponds to the local efficiency of star formation. This local efficiency is computed based on the magneto-thermodynamical properties of a parent cell and its close neighbours. We define this efficiency following the multi-freefall PN model from \citet{Federrath2012},
\begin{equation}
\epsilon_\text{ff} = \frac{\epsilon_\text{cts}}{2 \phi_t} \exp{\left(\frac{3}{8} \sigma_s^2\right)} \left[1 + \text{erf}\left(\frac{\sigma_s^2 - s_\text{crit}}{\sqrt{2 \sigma_s^2}}\right) \right].
\end{equation}
In this definition, $\epsilon_\text{cts}$, set to 0.5, represents the maximum amount of gas that can fall onto stars in the presence of proto-stellar feedback. $\sigma_s$ is the dispersion of the logarithm of the gas density to the mean gas density $s = \ln{\left(\rho / \left<\rho\right>\right)}$. $s_\text{crit}$ is the critical density above which post-shock gas in a magnetised cloud is allowed to collapse against magnetic support \citep{Hennebelle2011,Padoan2011}. It is defined as 
\begin{equation}
    s_\text{crit} = \ln{\left(0.067 \; \theta^{-2} \alpha_\text{vir} \mathcal{M}^2 f\left(\beta\right)\right)},
\end{equation}
with $\mathcal{M}$ being the Mach number and  $\alpha_\text{vir}$ the virial parameter, computed as indicated in \citet{Kimm2017}. $f \left( \beta \right)$ is a function of $\beta$ defined by Equation 31 in \citet{Padoan2011}. Finally, we set $\phi_t = 0.57$ and $\theta = 0.33$, with the values for these parameters extracted from \citet{Padoan2011} best fit values for multi-scale models of star formation in magnetised giant molecular cloud simulations.
This star formation prescription has already been introduced in various studies \citep{Trebitsch2017,Mitchell2018,Rosdahl2018} and in particular by \citet{Kimm2017}, and will be analysed in more detail in Devriendt~et~al.~(in preparation).

\begin{figure}
\centerline{\includegraphics[width=0.8\columnwidth]{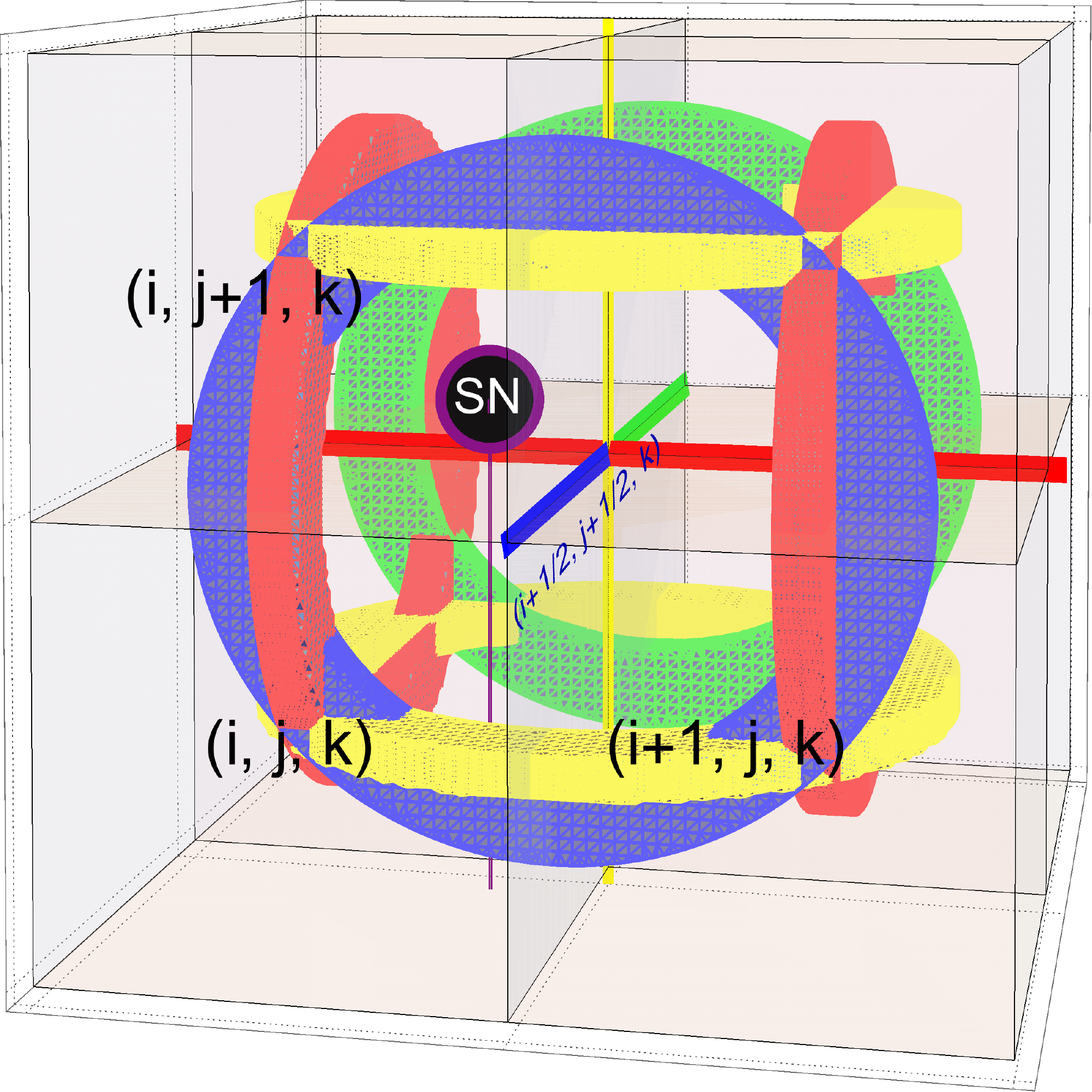}}
\caption{Illustration of the magnetic injection prescription employed during SN explosions on the neighbouring 8 cells. Coordinates in black show cell centred positions. The position of a supernova (SN) event occurring in cell $(i, j+1, k)$ is labelled by the black circle. The six rings represent the  magnetic loops injected in association with the SN. These rings increase the magnetic field of each cell face they traverse by $B_\text{inj}$, with the direction of the loop determined as explained in the text. Each ring is centred on a unique cell edge, to which it is associated, and which is depicted by a thick line of the same colour. In the text, we describe in more detail the injected loop around the cell edge $(i+1/2,  j+1/2, k)$, differentiated in blue.}
\label{LoopInjection}
\end{figure}

A fraction of the star particle mass is expected to explode via supernova \citep{Kroupa2001}.We take this into account by employing a model for supernova (SN) explosions presented in \citet{Kimm2014,Kimm2015}. For each SN, the model computes the amount of momentum that would be injected by the Sedov-Taylor blast wave solution dictated by the local spatial resolution of the simulation. This is related to the phase of the SN resolved by the simulation. Accordingly, each SN event injects mass and momentum into its host cell and the neighbours of that cell. Each SN occurs $3$ Myrs after the formation of its host star particle, and returns a mass fraction $\eta_\text{SN} = 0.213$ with a metallicity of $\eta_\text{Z} = 0.075$ back to the ISM. We assume an average mass of $19.135 M_\odot$ for a typical SN, consistent with a \cite{Kroupa2001} stellar IMF. Furthermore, each SN also injects back into its immediate 8 neighbouring cells a coiled, divergence-less magnetic field. We illustrate the method in Figure~\ref{LoopInjection}.  For a SN located in cell $(i, j+1, k)$, we identify the epicenter of the magnetic field injection at position $(i+1/2, j+1/2, k+1/2)$. As can be seen in Figure~\ref{LoopInjection}, each of the 6 cell edges emerging from this vertex is surrounded by 4 cells and is at the intersection of 4 cell faces. A closed loop defined by a line traversing these and only these 4 interfaces has an inherent null divergence. We take advantage of this fact to inject a constant magnetic field of $B_\text{inj} = 10^{-5}$~G into each of these cell faces, divergence-less by construction. As an example, for edge $(i+1/2,j+1/2,k)$, the magnetic fields of the 4 cell faces adjacent to this edge are modified following 
\begin{equation}
\begin{split}
B_x^{i+1/2,j+1,k} &= B_x^{i+1/2,j+1,k} \pm B_\text{inj} \, , \\
B_y^{i+1,j+1/2,k} &= B_y^{i+1,j+1/2,k} \mp B_\text{inj} \, , \\
B_x^{i+1/2,j,k} &= B_x^{i+1/2,j,k} \mp B_\text{inj} \, , \\
B_y^{i,j+1/2,k} &= B_y^{i,j+1/2,k} \pm B_\text{inj} \, . \\
\end{split}
\end{equation}
This process is repeated for the other 5 edges, modifying the faces with magnetic components perpendicular to the direction of the edge. The choice of sign for the injected field is selected to align with the majority of the local magnetic field, maximising the local injection energy. We note that regardless of the injection configuration, every compact divergence-less injection mechanism will in most cases oppose some part of the previously existing magnetic field. This implies that for a fixed $B_\text{inj}$, each group of 4 cells can increase its magnetic energy density in a range within $\epsilon_\text{mag} = 0.5 {-} 1.5 \times 10^{-10} \text{ergs}\ \text{cm}^{-3}$. Only in the case of the local magnetic field having the exact same configuration as the injection will the efficiency of the injection be maximised. In terms of energy, each SN injects $E^\text{mag}_\text{SN} \sim 10^{48} {-} 10^{49}$~ergs to the simulation, once again depending on the local configuration of the field and the volume of the injected cells. Subsequent expansion of the SN explosions will slowly erase the topological configuration of the injection. This magnetic injection procedure will be reviewed in more detail together with alternative SN injection mechanisms in Martin-Alvarez~et~al.~(in preparation). Magnetic field injections are added to the total and the SN-tracer magnetic fields.

In total, four sets of initial conditions were generated, varying only in the strength of the primordial B-field. In all simulations, this primordial magnetic field was set to a simple uniform value with direction along the $z$-axis. This primordial magnetic field is initialised for the total magnetic field and the primordial-tracer magnetic field. We note, however, that the configuration of the primordial magnetic field could have some potential impact on the results \citep[][although see \citealt{Dolag2002}]{Marinacci2015} and more sophisticated configurations should be explored. Our four simulations have corresponding comoving primordial magnetic fields $B_0 = 10^{-14} \text{G}$ (B14), $B_0 = 10^{-13} \text{G}$ (B13), $B_0 = 10^{-12} \text{G}$ (B12), and $B_0 = 10^{-11} \text{G}$ (B11). These strengths are well within aforementioned observational constraints of the present-day cosmic magnetic field, and lead to magnetic fields on the order of $\mu \text{G}$ in the simulated galaxies shortly after their collapse. Whenever used, haloes are identified in the simulations with the AMIGA Halo Finder \citep{Gill2004,Knollmann2009}, using the spherical top-hat collapse model to determine the over-density at a given redshift that will collapse.

\section{Results}
\label{s:Results}

Here we present the first results of our magnetic field tracer algorithm, reviewing its numerical accuracy and its potential to uncover the properties of magnetic fields with different origins.  The goal of this work is to demonstrate the capabilities of our new MHD tracer algorithm and here we focus on separating the magnetic contribution from a primordial magnetic field, and the magnetic field injected back into the gas when stars explode. Each of these fields operates in different locations and in this first work, we focus on the difference between these two contributions.  All four simulations with different primordial magnetic seed strengths are evolved until $z=6$ and in what follows, we study the magnetic fields from the largest scales resolved by our simulation to the smallest scales in the ISM of galaxies. 

\subsection{Tracing the topology of the large scale magnetic field}
In the top panel of Figure~\ref{MHDrgb}, we show the redshift evolution of the large scale magnetic field from $z=8-6$.  The primordial magnetic field, shown in red, traces the large scale filamentary structure of the matter distribution while blue regions, representing the SN-injected magnetic field, emanate from the most massive galaxies and penetrate  into the IGM perpendicular to the filaments.  At high redshift, the magnetic field is initialised in the $z$-direction with constant magnitude.  As the Universe expands, the energy density of the primordial field decreases in the IGM while being increased as gas condenses into filaments and galaxies. This process will continue as filaments accrete gas and become denser and more massive. However, towards $z=6$, the strength of the UV background drastically increases \citep{Bolton2007,Calverley2011,Wyithe2011}. This is modelled as a uniform heating term in our simulation and has the effect of evaporating the filaments that are not self-shielded \citep{Pawlik2009}. Comparing the snapshot at $z=7$ with $z=6$, the most diffuse filaments that appear red at $z=7$ are missing at $z=6$ as the UV background has efficiently reduced their density, thus reducing the magnetic field strength. By $z = 6$, the vast ensemble of filaments have yet to be significantly impacted by the SN-injected magnetic fields, retaining memory of the configuration of primordial magnetic fields in the simulation.

While the large-scale filamentary structure of the universe is beginning to take shape, the first generation of stars begin to form in the earliest collapsing objects and the total magnetic energy in the SN-injected component increases with decreasing redshift, as more SNe occur. Repeated star bursts allow for the magnetic energy to build up around galaxies from the inside-out. SN winds push magnetised gas out of the halo and into the IGM where it then expands, leading an adiabatic decrease of the magnetic field. 

The bottom panel of Figure~\ref{MHDrgb} shows the magnetic field around the three most massive galaxies at $z=6$. In the left column, three dense filaments are feeding a halo with $M_{\rm vir}=10^9{\rm M_{\odot}}h^{-1}$. These filaments appear bright red in the image and the intensity increases towards the galaxy as magnetic fields are frozen in the gas in ideal MHD and this gas becomes denser. The magnetic field injected from SNe is clearly visible in this image, emanating from the galaxy into the low density regions between the filaments. At the centre of the halo, the red and blue colours appear blended as there is a contribution from both the primordial field, and the SN-injected field in the same spatial location. When these fields have similar orientation, the total field is amplified; however, in certain regions, the SN-injected field opposes the primordial field which reduces the overall magnetisation in the region. This effect is further explored in Section~\ref{ss:GlobalProp}.

The second most massive halo appears very similar to the first, as dense, red filaments are feeding the galaxy, while blue magnetised gas emanates from the central regions of the system. In contrast, the third most massive system has a magnetic environment that is completely dominated by the SN-injected component as the filaments feeding the galaxy have likely been disrupted by a combination of an increasing UV background, and SN feedback. The environment around this galaxy is almost entirely blue. Because this region is less dense compared to the other two, the magnetic winds driven by SN more easily penetrate into the low density regions of the IGM.

\begin{figure}
\centerline{\includegraphics[scale=1]{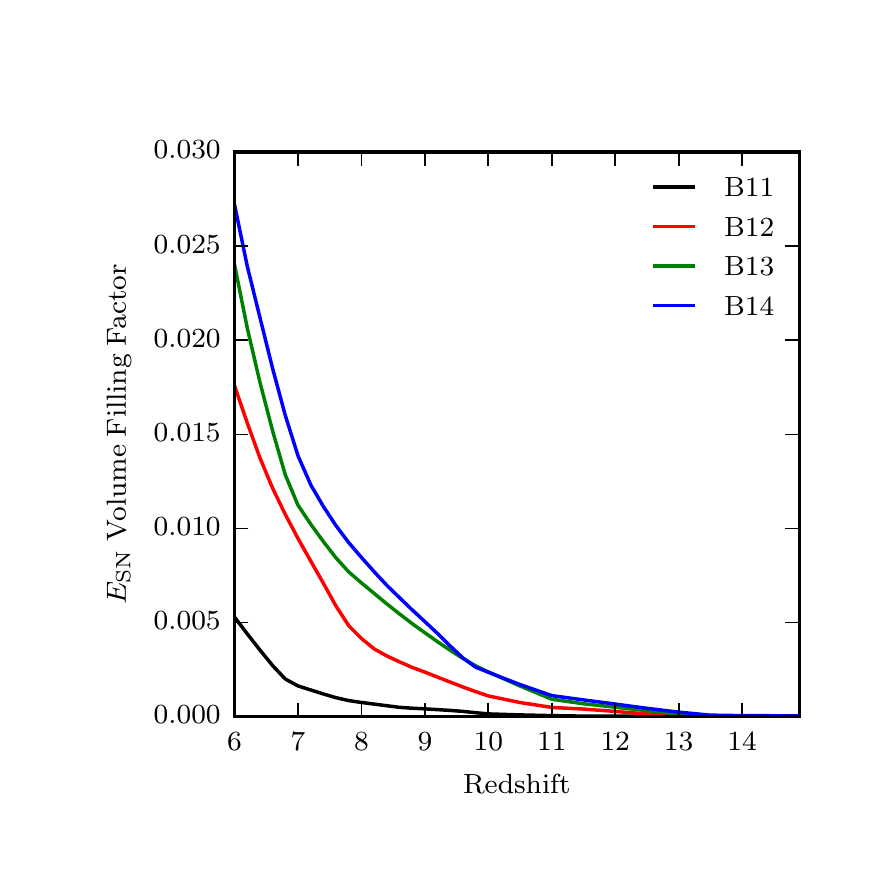}}
\caption{Volume filling factor of the regions where at least 1\% of the total magnetic energy in the cell has come from SN explosions as a function of redshift. Different colour-codings indicate the runs with different initial magnetic field strength.  Even though the SN magnetic energy dominates the total magnetic energy in the B13 and B14 simulations, less than 3\% of the total volume is significantly affected by this component of the magnetic field.}
\label{vff}
\end{figure}

\begin{figure*}
\centerline{\includegraphics[scale=1]{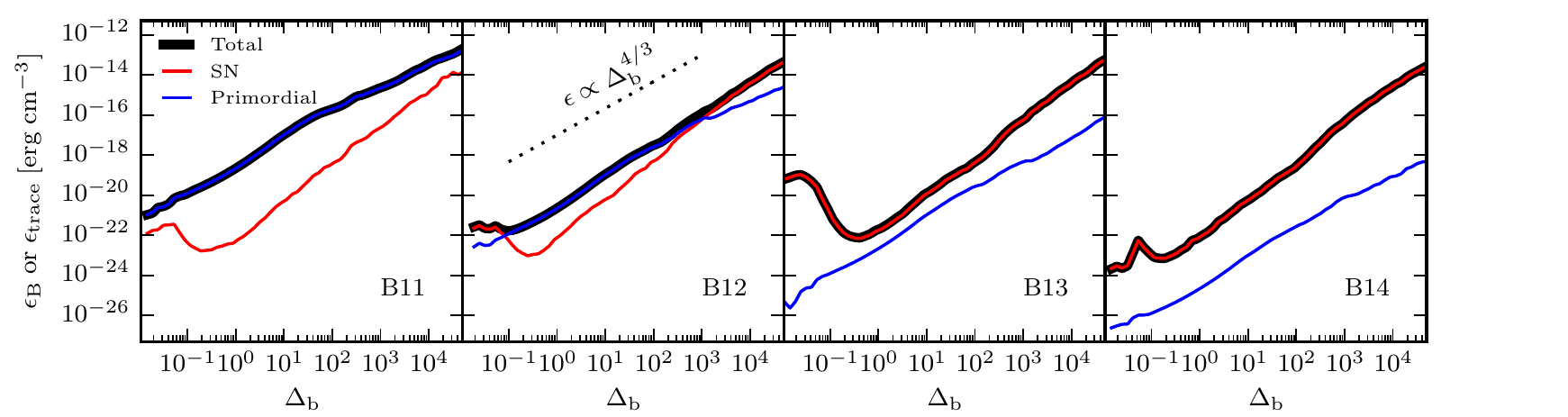}}
\caption{Total magnetic energy density (summed over all cells in the simulation) and magnetic energy density in each of the tracer components as a function of baryonic over-density at $z=6$ for each simulation.  For low values of the primordial magnetic field, the SN field dominates the total energy over the whole box at all over-densities while the reverse is true of high values of the primordial magnetic field. When the total magnetic energy is in approximate equipartition (B12), SN magnetic energy dominates at the highest and lowest over-densities (i.e. inside galaxies and SN heated regions) while the primordial magnetic field dominates at mean density (i.e. the IGM). }
\label{enod}
\end{figure*}

\subsection{Global Properties}
\label{ss:GlobalProp}
The energy contained in the tracer fields ($E_{\rm trace}$) within a given cell is:
\begin{equation}
    E_{\rm trace} = \frac{1}{2}V_{\rm cell}\left[\vec{B}^2_{\rm SN}+\vec{B}^2_{\rm Primordial}+2 \left(\vec{B}_{\rm SN} \cdot \vec{B}_{\rm Primordial}\right)\right],
\label{eq:CrossTerm}
\end{equation}
where $V_{\rm cell}$ is the volume of the cell, $B_{\rm SN}$ is the magnetic field in the SN-injected component, and $B_{\rm Primordial}$ is the magnetic field contained in the primordial component.  The key aspect of this equation is that by splitting the total magnetic field into multiple components, cross terms between these components appear in the energy calculation. These terms are not positive in the case where magnetic tracer fields are oriented in opposite directions.

In Figure~\ref{enfrac}, we show the fraction of the total energy contained in the primordial field, the SN-injected field, and the cross term as a function of redshift for each of the four simulations. In the B14 simulation (right panel), which contains the weakest primordial magnetic field, as soon as the first generation of stars explode at $z\sim16$, the total energy contained within the SN-injected field is nearly equal to the total magnetic energy in the primordial field. As these SN bubbles expand, the energy in the SN-injected magnetic field quickly dissipates. However, by $z=14$ much more sustained star formation occurs in the simulation and the energy in the SN-injected magnetic field quickly dominates that of the primordial field such that in the redshift range $13\geq z\geq6$, the total magnetic energy in the simulation is completely dominated by that from SN injections. 

By increasing the strength of the primordial magnetic field, the redshift at which the SN-injected field dominates the total magnetic energy occurs later. For the B13 and B12 simulations, equipartition occurs at $z\sim10$ and $z\sim6$, respectively, while for the B11 simulation, the energy in the primordial field completely dominates the total magnetic energy in the box at all redshifts simulated.

The cyan lines in each of the panels of Figure~\ref{enfrac} represent the magnetic energy contained in the cross term. Interestingly, regardless of the simulation, the energy in this component rarely reaches more than a few percent of the total. This component can be both positive and negative depending on the exact orientation of the tracer fields near galaxies; however, in neither case does it ever represent a significant fraction of the total in our simulations. This is to be expected, as the relevance of this term severely depends on the correlation between the magnetic fields being traced: the cross term between two tracers becomes most important when the two magnetic fields are comparable. For the ones employed in this manuscript, one tracer typically dominates over the other. However, primordial and SN-generated magnetic fields can develop a non-negligible cross term energy component at later stages in the evolution of galaxies, once other mechanisms like their large-scale rotation become important.

Even though the energy contained in the SN-injected magnetic field dominates the total magnetic energy by $z=6$ in both the B13 and B14 simulations, this does not necessarily mean that the majority of the volume of the simulation is affected by this magnetic component. Based on the images shown in the top row of Figure~\ref{MHDrgb}, the regions magnetised by SN-injections rarely extend more than $\sim100$~kpc from their host galaxies and most of the volume of the simulation is only aware of the presence of the primordial magnetic fields. We quantify this in Figure~\ref{vff} by plotting the volume filling factor of the energy contained in the SN-injected magnetic field as a function of redshift for each of the four simulations. We define this quantity to be the fraction of the simulation volume where at least 1\% of the local magnetic energy is comprised from the SN-injected component. This quantity may give insight into where in the Universe to look in order to detect the effects of primordial magnetic fields.  By $z=6$, the volume filling factor of SN-injected magnetic energy is only $\sim2.7\%$ for the B14 simulation, decreasing to $\sim0.5\%$ for the B11 simulation. Thus in all simulations, the volume of the simulation filled by SN-injected magnetic energy is essentially negligible compared to the total. Primordial magnetic fields are expected to dominate a majority of the cosmic volume, even in the presence of strong astrophysical sources \citep{Vazza2017}.

As redshift decreases, the rate at which SN-injected magnetic energy is filling the volume is accelerating. Thus if we were to run this simulation for another $\sim13$Gyr to $z=0$, it is likely that SN injected fields may be able to fill a significant portion of the IGM. The exact volume filling fraction however is very subject to how star formation and SN injections are modelled and this is further discussed in Section~\ref{s:Caveats}.  The star formation rate density is expected to turn over at $z\sim2$ \citep{Madau2014} and thus the volume filling factor cannot grow indefinitely.  Nevertheless, magnetised winds emanating from galaxies may provide a plausible channel for magnetising the IGM by $z=0$ \citep[e.g.][]{Dubois2010,Beck2013,Vazza2017}. 

\begin{figure*}
\centerline{\includegraphics[scale=1]{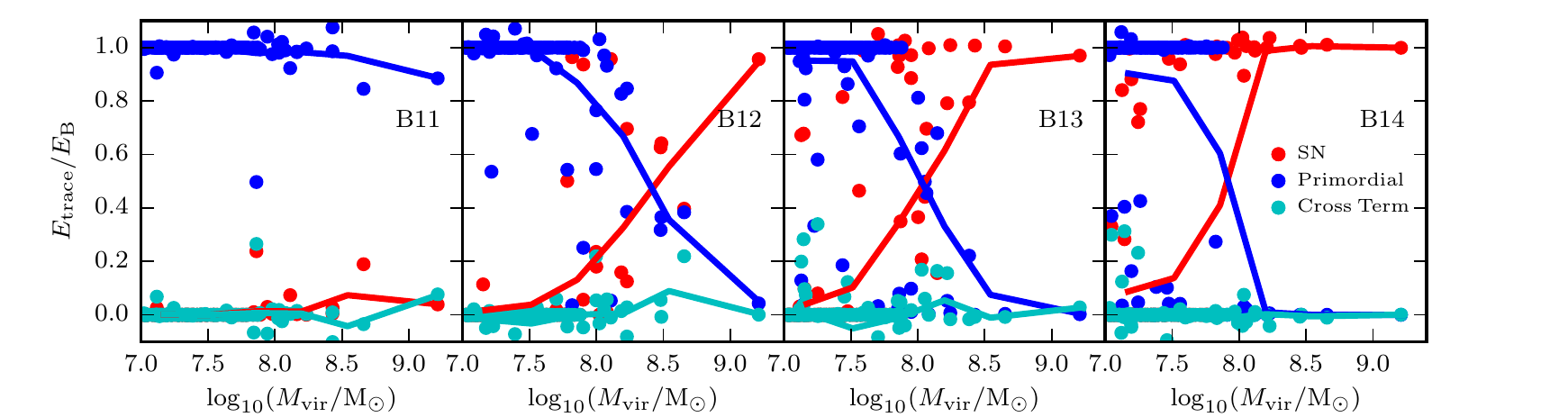}}
\caption{Ratio of the fraction of total magnetic energy in each of the tracer magnetic fields compared to the total within the virial radius of a halo as a function of virial mass for each of the four simulations at $z=6$.  Red, blue, and cyan represent the energy in the SN injected field, the primordial field, and the cross term, respectively.  The coloured lines show the average ratio in bins of virial mass.  The virial mass where equipartition occurs between the SN injected and primordial fields increases with the strength of the primordial magnetic field.  Equipartition never occurs in the B11 simulation for any halo mass probed by our simulation.}
\label{erh}
\end{figure*}

\begin{figure*}
\centerline{\includegraphics[scale=1]{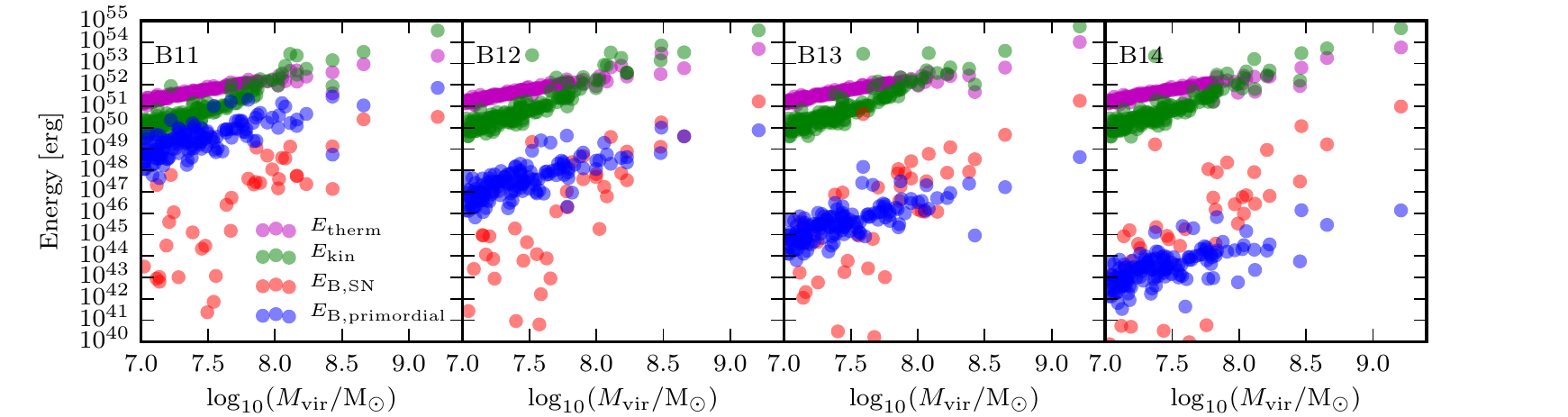}}
\caption{Total thermal energy (magenta), kinetic energy (green), magnetic energy in the SN injected field (red), and magnetic energy in the primordial field (blue) within the virial radius of individual haloes in each of the four simulations at $z=6$.  The thermal energy dominates for low mass haloes while the kinetic energy is dominant for haloes with $\log_{10}(M_{\rm vir}/{\rm M_{\odot}}) > 8.0$.  The total primordial magnetic energy within haloes decreases with the strength of the primordial magnetic field.  In no simulation is the magnetic energy in the entire halo in equipartition with its thermal or kinetic energy.}
\label{esh}
\end{figure*}

Although the majority of the volume of the simulation is not affected by the SN-injected magnetic field, one of the main advantages of the MHD tracer algorithm is that we can separate the affects of each tracer field both in time (as we have shown) as well as based on environment. In Figure~\ref{enod}, we plot the fraction of total magnetic energy density at a given gas density, $\Delta_{\rm b}=\rho_{\rm b}/\bar{\rho}_{\rm b}$ where $\bar{\rho}_{\rm b}$ is the mean baryon density at a given redshift, for the four simulations at $z=6$.  The total magnetic energy density at fixed density is well described by a power-law in all simulations, except at $\Delta_{\rm b}\lesssim 0.1$ where SN dominate the energy density in the B12, B13, and B14 simulations and strong deviations from the mean trend can be seen.  This density represents a very small fraction of the total volume of the simulation and can rapidly change with time depending when the most recent SN occurred.

The blue lines in each panel of Figure~\ref{enod} represent the energy density in the primordial magnetic field as a function of density. For adiabatic compression, the magnetic energy density is expected to scale as $\Delta_{\rm b}^{4/3}$. Blue lines exhibit power-law slopes consistent with this value. The red lines exhibit steeper slopes than $\Delta_{\rm b}^{4/3}$ because they are injected at a higher strength than the local primordial field at a fixed density and then they cascade to lower densities as the SN bubbles expand (see \citealt{Vazza2017} who also find a steeper slope for the injected field). The B13 and B14 simulations (third and fourth panels) are dominated by SN-injected magnetic energy at all densities and thus the global line has a steeper slope than $4/3$. In contrast, the B11 (first panel) simulation is dominated by the primordial magnetic field at all densities and thus exhibits a slope of $4/3$. The B12 simulation (second panel), which exhibits equipartition in total magnetic energy at $z=6$ between the two tracer fields, shows the most interesting behaviour in the magnetic energy density as a function of gas density.  At $\Delta_{\rm b}\gtrsim1000$ and $\Delta_{\rm b}\lesssim0.1$ the SN-injected energy density dominates while at densities between these two regimes, the primordial component dominates. Thus in this simulation, although there is energy equipartition between the two MHD tracer fields at $z=6$, the SN-injected field only dominates well inside of galaxies and in SN remnants.

\subsection{Magnetic fields in haloes}
In the previous section, we analysed the global properties of each component of the magnetic field across the entire simulation volume. In this section, we study the properties of each component of the magnetic field inside of virialised haloes.

\begin{figure*}
\centerline{\includegraphics[scale=1]{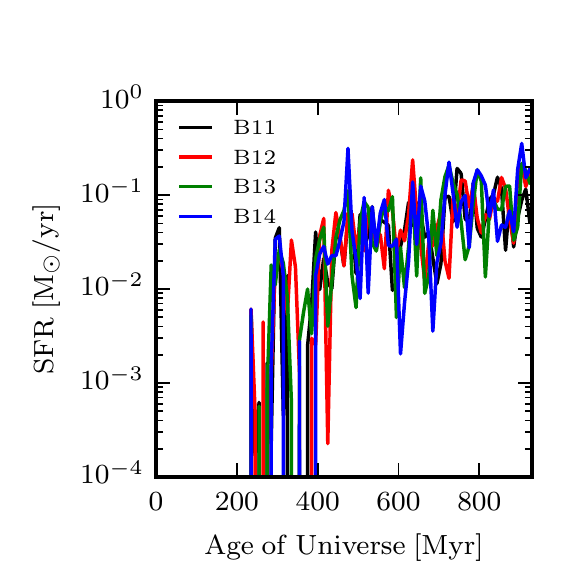},\includegraphics[scale=1]{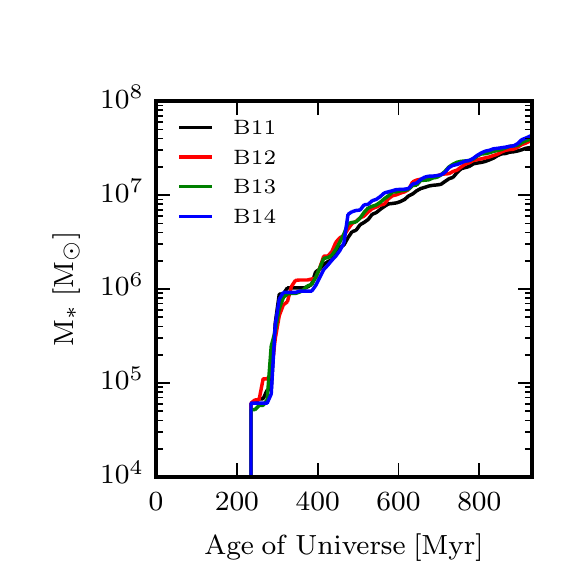}\includegraphics[scale=1]{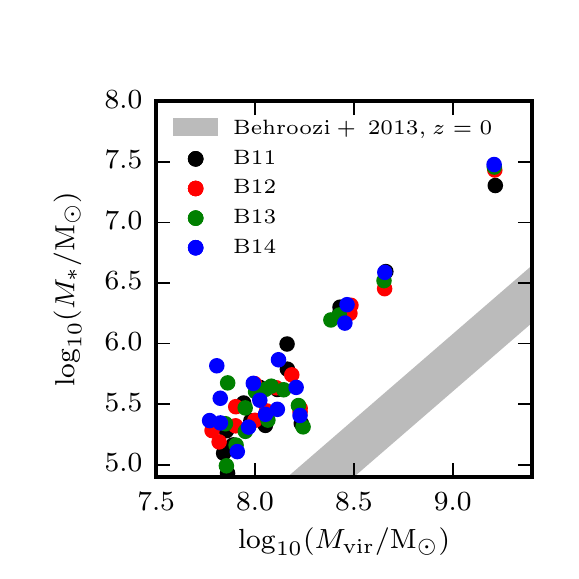}}
\caption{(Left) Star formation rate as a function of the age of the Universe for each simulation.  (Centre) Total stellar mass as a function of the age of the Universe.  (Right) Stellar mass-halo mass relation for central galaxies in each of the four simulations.  The grey shaded region shows the local extrapolated stellar mass-halo mass relation from \protect\cite{Behroozi2013}. }
\label{smass}
\end{figure*}

The first question we aim to address is what percentage of the total magnetic energy within the virial radius of haloes is represented by the primordial and SN-injected fields. In Figure~\ref{erh} we plot the ratio of total energy in each tracer field to the total energy within a halo as a function of halo mass at $z=6$. In all simulations most of the haloes with $M_{\rm vir}\lesssim10^8{\rm M_{\odot}}$ have magnetic energy completely dominated by the primordial component. These systems are extremely inefficient at forming stars \citep[e.g.][]{Kimm2017} and thus the amount of magnetic energy injected during SN is limited in these galaxies. However, for more massive haloes in the B12, B13, and B14 simulations, the SN-injected magnetic energy becomes very important and at $M_{\rm vir}\sim10^9{\rm M_{\odot}}$, most of the energy contained within the haloes is dominated by the SN-injected component. The mass at which the crossover occurs between SN-injected dominated and primordial-dominated shifts to progressively larger mass for increasing primordial magnetic field strength. For the B12 and B13 simulations, there is considerable scatter at intermediate masses between which tracer field dominates the total energy and this is due to the exact star formation history for an individual halo. In no simulation does the cross term contribute on average a significant fraction of the total energy, consistent with the results of Figure~\ref{enfrac}. In certain haloes, the cross term can contribute up to $\sim40\%$ of the total energy; however, these cases are rare.  SKA will have the potential of observing magnetic fields in the surroundings of galaxies and clusters, which could probe primordial magnetic fields \citep{JH15,Taylor15}. Our results allow us to constrain in our simulations a maximal halo mass below which, for a given primordial magnetic field strength, the magnetic energy budget of a halo is dominated by the primordial component. Upcoming applications of this algorithm have the potential to determine the regions around galaxies where future observations should aim to detect magnetic fields of primordial nature, and which halo masses are still dominated by primordial magnetic fields.

A natural question is whether the energy contained within the magnetic field in a halo is in equipartition with the thermal and kinetic energy of that halo. In Figure~\ref{esh}, we plot the total thermal energy, kinetic energy, and energies in each of the MHD tracer fields as a function of halo mass at $z=6$ for each of the four simulations. In no simulation is the total magnetic energy comparable with either the thermal or kinetic energy. At the grid resolutions obtained in our dark matter haloes, turbulent amplification in the halo is expected to be negligible if at all existent, compared with the magnetic energies expelled from the galaxies. Typical morphologies of observed magnetic fields in haloes are also consistent with being driven by galactic outflows \citep{Dahlem97}. For the B11 simulation, with the highest primordial magnetic field, only for the lowest mass galaxies does the total magnetic energy approach values of $10\%$ of the kinetic energy. For most other mass systems, the magnetic energy is far below equipartition.

\subsection{The effects of magnetic fields on star formation}
Because of the variations in total strength of the magnetic fields within haloes of different mass (see Figure~\ref{esh}), it is interesting to understand whether the magnetic fields have an impact on star formation in galaxies.  In the left and central panels of Figure~\ref{smass}, we show the star formation rate (SFR) and total stellar mass formed as a function of the age of the universe for each of the four simulations with varying primordial seed strengths. In all simulations, by $z=6$, the total mass in stars formed is nearly identical as are the SFRs. Stochasticity in our star formation algorithm means that at a fixed time, the SFR is expected to deviate between the simulations by a small fraction; however, it is clear that the general trend is the same between all four simulations.

Since the relationship between stellar mass and halo mass is expected to be reasonably steep such that more massive haloes are expected to form stars much more efficiently \citep{Moster2013,Behroozi2013}, total stellar mass formed and total SFR in the simulation may only be representing the high mass haloes in the simulation. Thus in the right panel of Figure~\ref{smass}, we plot the stellar mass-halo mass relation at $z=6$ for each of the four simulations.  We find no systematic offsets in stellar mass for a given halo mass between the simulations, even for the lowest mass haloes, indicating that the magnetic fields in our simulation have a negligible effect on the resulting stellar masses, regardless of the primordial seed strengths that we have chosen to model.  This is true even for the B11 simulation which has a magnetic field strength at the highest gas densities that is larger than the strengths that our SN-injection can obtain (see Figure~\ref{enod}). Note that we over-predict the expected stellar mass for a given mass halo compared to predictions from abundance matching \citep{Behroozi2013}.  No change in the global SFR may be expected because our primordial seed strengths are not high enough to prevent accretion of gas onto haloes. For this one needs a primordial seed strength of $\sim10^{-9}$~G \citep{Marinacci2016}.  

\section{Caveats}
\label{s:Caveats}
Our algorithm is an effective tool to better understand the amplification and evolution of magnetic fields with different origins.  All of the limitations that come with numerical MHD are also found in our simulations. Astrophysical MHD simulations have numerical viscosities and resistivities significantly higher than their physical counterparts in nature, which limits our capabilities to accurately model various processes of critical importance (e.g. small-scale dynamo, grid-driven diffusion, turbulent cascade).   Note that a turbulent dynamo is expected to be capable of rapidly amplifying primordial magnetic fields within galaxies in timescales as low as $\tau \sim 25 - 300\; \text{Myr}$ \citep{Schober2013}. This is not seen in our current simulations.  With the primordial magnetic field serving as a seed for this dynamo, this process could naturally increase the importance of primordial magnetic fields in haloes in our simulation. Our simulations do resolve the amplification of magnetic fields in compact objects. Thus the strength of the magnetic field that we inject, how often we inject (based on uncertain SN rates), and the properties of the injected fields are assumptions. Furthermore, the feedback in our simulation is not strong enough to regulate star formation so we may over-predict the amount of injected magnetic energy. Nevertheless, we intend our simulations to be a demonstration of the new algorithm rather than a complete physical model.

\section{Conclusions}
\label{s:Conclusions}

The goal of this manuscript is to present a first demonstration of the MHD tracer algorithm. We presented the mathematical foundation for the employed decomposition of the physical magnetic fields in the simulations in Section~\ref{s:Algorithm} and then described our implementation in the CT MHD code {\sc RAMSES}. We performed four different high-resolution cosmological MHD simulations with {\sc RAMSES} using different strengths of a primordial magnetic field combined with an additional magnetic component that is injected when SN explode. We used these simulations to showcase the accuracy of our new algorithm (see Appendix~\ref{appB}) and demonstrate the first results regarding the importance of primordial magnetic fields versus supernova injected magnetic fields in small cosmological volumes at high redshift. The main conclusions of this work are:

\begin{itemize}
    \item The equations for ideal MHD can be linearly decomposed so that individual contributions to the total magnetic field from different sources can be tracked.
    \item Our method conserves total energy, total B-field on each cell face, and maintains the solenoidal constraint so that it is neither dynamically important nor affecting the induction equation.
    \item The dominant component of total magnetic energy depends on the strength of primordial magnetic fields (as well as how much is injected during each SN). The relevance of the primordial magnetic field is expected to decrease with time due to cosmic expansion unless it can be significantly amplified in haloes.  In contrast, SN-injected fields become more important with cosmic time as more stars are formed.
    \item Primordial magnetic fields and those injected from SN possess different topologies in the context of the large-scale structure. Primordial magnetic fields are found to dominate the majority of the volume of the simulations, whereas SN-injected magnetic fields are confined to galaxies and the vicinity of their dark matter haloes by $z=6$.  The volume filling factor of SN-injected magnetic fields are expected to increase as a function of decreasing redshift.
    \item Within dark matter haloes, there exists a lower limit on the virial mass ($M_\text{vir} \sim 10^{8}-10^{8.5} M_\odot$) below which SN-injected magnetic fields lose relevance compared to primordial magnetic fields. This mass limit is found to depend on the strength of the primordial magnetic field, and is also expected to be dependent on numerical resolution and the filtering mass due to reionization.
    \item Consistent with other work \citep{Marinacci2016}, we find the global cosmic star formation properties to be unchanged within the range of primordial magnetic fields strengths probed in this work.
\end{itemize}

These results are subject to the caveats discussed in Section~\ref{s:Caveats}. Nevertheless, they represent the first demonstration of separating the contribution of primordial magnetic fields versus SN-injected fields in the same simulation, providing a methodology for future, more detailed studies of magnetogenesis. Future work will encompass a more comprehensive list of sources of magnetic fields (e.g. AGN, turbulence, cosmic rays) as well as study different seeding mechanisms such as in intergalactic ionisation fronts or various primordial fields that may be generated by inflation.

\section*{Acknowledgements}
We thank the anonymous referee for their comments that improved the manuscript.  This work made considerable use of the open source analysis software {\small PYNBODY} \citep{pynbody}. HK's thanks Brasenose College and the support of the Nicholas Kurti Junior Fellowship as well as the Beecroft Fellowship. This work was supported by the Oxford Centre for Astrophysical Surveys which is funded through generous support from the Hintze Family Charitable Foundation. TK acknowledges support by the National Research Foundation of Korea to the Center for Galaxy Evolution Research (No. 2017R1A5A1070354) and in part by the Yonsei University Future-leading Research Initiative of 2017 (RMS2-2017-22-0150).  

This work used the DiRAC Data Centric system at Durham University, operated by the Institute for Computational Cosmology on behalf of the STFR DiRAC HPC Facility (www.dirac.ac.uk).  This equipment was funded by the BIS National E-infrastructure capital grant ST/K00042X/1, STFC capital grant ST/K00087X/1, DiRAC operations grant ST/K003267/1 and Durham University.  DiRAC is part of the National E-Infrastructure.

\appendix
\section{Energy and magnetic field conservation}
\label{appB}

\begin{figure*}
\centerline{\includegraphics[scale=1]{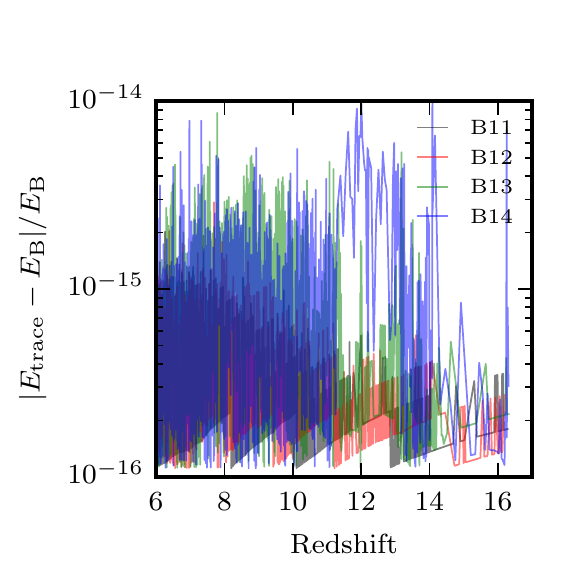},\includegraphics[scale=1]{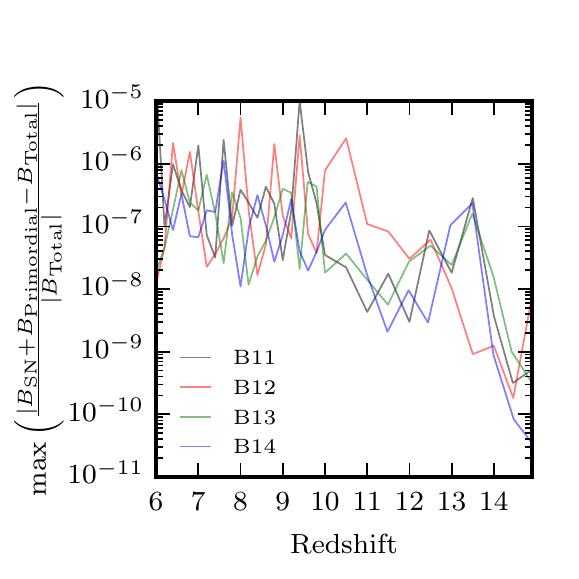}\includegraphics[scale=1]{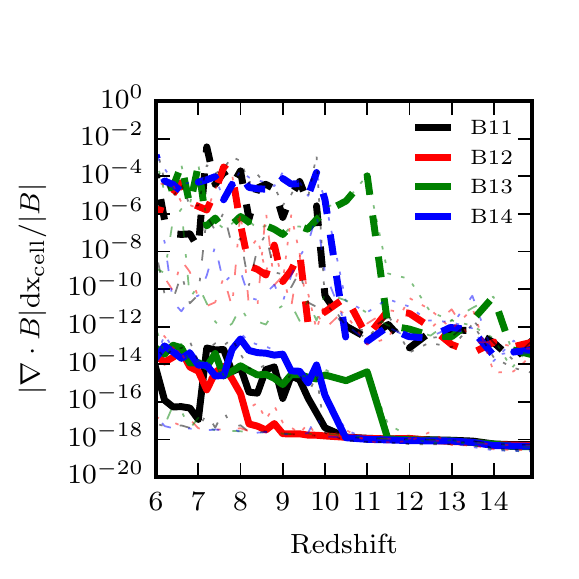}}
\caption{(Left) Fractional energy difference ($|E_{\rm trace}-E_{\rm B}|/E_{\rm B}$) between the sum of the energies stored in the tracer components and the total magnetic energy in the box as a function of redshift.  The total magnetic energy is conserved to nearly machine precision regardless of the strength of the primordial magnetic field.  (Centre) Maximum difference between the total magnetic field and the sum of the tracer fields (${\rm max}\left(\frac{|B_{\rm SN}+B_{\rm primordial}-B_{\rm total}|}{|B_{\rm total}|}\right)$) in all cell faces and all directions in the simulation as a function of redshift. The maximum difference between the magnetic field stored in the tracer components never deviates more than 0.001\% for any cell face over the course of the simulation.  (Right) Average (solid lines) and maximum (dashed lines) divergence with respect to the local B-field ($|\del\cdot B|{\rm dx_{cell}}/|B|$) in the total magnetic field.  The thin dashed and dot-dashed lines show the same quantities for the SN and primordial magnetic field respectively.  Neither the total, nor the individual tracer fields exhibit divergences that represent more than 1\% of the total magnetic field at any point in the simulation.  On average the divergences remain below $10^{-13}$ of the local B-field.}
\label{cons}
\end{figure*}

In this Appendix, we demonstrate the robustness of our algorithm by directly measuring the conservation of total magnetic energy, the magnetic field on the faces of individual cells, and the solenoidal constraint.

In the left panel of Figure~\ref{cons} we plot the fractional difference between the sum of the total energy contained within each of the tracer fields and the total magnetic field in each of the four simulations (B11-B14). The total energy contained in the tracers is defined by Equation~(\ref{eq:CrossTerm}). The total energy within the simulation volume is conserved nearly precisely with a fractional difference that does not exceed $10^{-14}$ throughout the course of the simulation. This value fluctuates when magnetic energy is injected with SN explosions; however, it is clear that the energy within the total magnetic field is well-captured even if it is split into multiple components.

In the central panel of Figure~\ref{cons}, we plot the maximum difference between the magnetic field contained within the tracers and the total magnetic field on all cell faces within the simulation for all four simulations. We divide this difference by the average local magnetic field strength across all six cell faces to avoid local X- and O-points. Over the course of the simulation, the deviation in the total magnetic field contained in the tracers from the total magnetic field in any of the faces never reaches more than $10^{-5}$ of the local magnetic field strength. The maximum error does increase sightly with time before tapering off as more SN injections occur; however, these errors are negligible compared to the local field strength. Hence the algorithm conserves the local magnetic field strength in addition to the total energy to high precision.

As shown in Section~\ref{s:Algorithm}, each individual tracer field must satisfy the solenoidal constraint. Because we use constrained transport, for both the total and each tracer field, the divergence should be constrained to near machine precision. In the third panel of Figure~\ref{cons}, we plot the average and maximum divergence in the total magnetic field as well as in each individual tracer field as a function of redshift. The average divergence in the simulation never increases to more than $10^{-12}$ of the local $B$-field and is thus well controlled by the constrained transport algorithm. The same is true for each of the tracer fields. With this precision, the divergence is neither dynamically important nor affecting the induction equation. See \cite{Hopkins2016} for a comparison of MHD schemes that use ``divergence-cleaning", where the divergence errors are considerably higher. Also shown in the right panel of Figure~\ref{cons} are the maximum divergences with respect to the local magnetic field as a function of redshift for each simulation. This value rarely surpasses $10^{-4}$ once again indicating that, even in the least controlled environments, the divergence is both negligible for the dynamics and the induction equation. Based on the three convergence tests described in this section, we have shown that the MHD tracer algorithm conserves all relevant quantities necessary for numerically modelling ideal MHD.

%%%%%%%%%%%%%%%%%%%%%%%%%%%%%%%%%%%%%%%%%%%%%%%%%%
\bibliographystyle{mnras}
\bibliography{main} % if your bibtex file is called example.bib
%%%%%%%%%%%%%%%%%%%%%%%%%%%%%%%%%%%%%%%%%%%%%%%%%%

% Don't change these lines
\bsp	% typesetting comment
\label{lastpage}
\end{document}